\documentclass[preprint]{elsarticle}

\usepackage{lineno,hyperref,amssymb}
\usepackage{graphicx}
\usepackage{amsmath}
\usepackage{xcolor}
\usepackage{tikz}
\usepackage[printwatermark]{xwatermark}

\newsavebox\mybox
\savebox\mybox{\tikz[color=black,opacity=0.2]\node{DRAFT};}

\modulolinenumbers[5]

\newcommand{\slaDelta}{\Delta \!\!\!\! / \;}
\newcommand{\slap}{p \!\!\! / \;}
\newcommand{\slaq}{q \!\!\! / \;}

\begin{document}
\allowdisplaybreaks

\begin{frontmatter}

  \title{
    \vskip-3cm{\baselineskip14pt
      \centerline{\normalsize DESY 20-156\hfill 0418-9833}
      \centerline{\normalsize September 2020\hfill}}
    \vskip0.5cm
    Moments $n=2$ and $n=3$ of the Wilson twist-two operators
    at three loops in the RI${}'$/SMOM scheme
  }




      \author[mymainaddress]{Bernd A. Kniehl}
      \ead{kniehl@desy.de}
      \address[mymainaddress]{II.~Institut f\"ur Theoretische Physik, Universit\"at Hamburg,\\
        Luruper Chaussee 149, 22761 Hamburg, Germany}
      \author[mymainaddress]{Oleg L. Veretin}
      \ead{oleg.veretin@desy.de}
      \address[mymainaddress]{Institut f\"ur Theoretische Physik, Universit\"at Regensburg,\\
        Universit\"atsstra\ss{}e 31, 93040 Regensburg, Germany}

      \begin{abstract}
We study the renormalization of the matrix elements of the twist-two
non-singlet bilinear quark operators, contributing to the $n=2$ and $n=3$
moments of the structure functions, at next-to-next-to-next-to-leading order in
QCD perturbation theory at the symmetric subtraction point.
This allows us to obtain conversion factors between the $\overline{\rm MS}$
scheme and the regularization-invariant symmetric momentum subtraction
(RI/SMOM, RI${}'$/SMOM) schemes.
The obtained results can be used to reduce errors in determinations of moments
of structure functions from lattice QCD simulations.
The results are given in Landau gauge.
      \end{abstract}


      \end{frontmatter}

\section{Introduction}

The great success of QCD in the description of the structure of hadrons relies
on the principle of factorization.
Phenomenologically, it is possible to access this problem only under some
particular kinematical conditions, as provided, for instance, in experiments
like deep-inelastic scattering, vector boson or heavy-meson production,
Drell-Yan process and others.

In hard processes, QCD factorization and scaling violation manifest themselves
in the well-known Dokshitzer-Gribov-Lipatov-Altarelli-Parisi (DGLAP) equation
\cite{Gribov:1972ri,Altarelli:1977zs,Dokshitzer:1977sg} and allow for
nonperturbative information, on how the parton momenta are distributed inside
the hadrons and how the hadron spins are generated, to be accumulated in parton
distribution functions (PDFs).
Besides PDFs, also other nonperturbative distributions and concepts like, e.g.,
light-cone distribution amplitudes (LCDAs) 
\cite{Chernyak:1977as,Efremov:1978rn,Lepage:1979za,Lepage:1979zb,Efremov:1979qk,Lepage:1980fj}
and generalized parton distributions (GPDs) \cite{Dittes:1988xz,Mueller:1998fv}
have been introduced.

At the operator level, the most significant contributions in hard processes
arise from operators of twist two.
In particular, in the case of non-siglet distributions,
bilinear quark operators play a crucial r{\^o}le.
Such operators, contributing to the $n$th moment of a distribution, are given
by symmetric traceless combinations, like
\begin{align}
\label{eq:Odef}
   {\cal S} \bar{\psi} \gamma_{\mu_1} D_{\mu_2} \dots D_{\mu_n} \psi \,,
\end{align} 
where the symbol $\cal S$ denotes total symmetrization over indices
$\mu_1,\ldots,\mu_n$ (including the factor $1/n!$) and subtraction of
all possible traces over pairs of indices.

Since the matrix elements of the operators in Eq.~(\ref{eq:Odef}) are of
nonpertubative nature, they can be accessed only by experiments, QCD sum rules,
or lattice-QCD simulations.
The most important examples of recent lattice studies include determinations of
low moments of LCDAs of mesons
(see, e.g., Refs.~\cite{Arthur:2010xf,Braun:2015axa,Braun:2016wnx,Bali:2019dqc})
and low moments of the proton PDFs and GPDs (see, e.g.,
Refs.~\cite{Aoki:2010xg,Green:2012ud,Bali:2018zgl,Harris:2019bih,Alexandrou:2020sml}).

To renormalize the matrix elements of the operators in Eq.~(\ref{eq:Odef}) on
the lattice, the regularization-invariant momentum-subtraction (RI/MOM) scheme
and its modification, the RI${}'$/MOM scheme, have been developed
\cite{Martinelli:1994ty,Franco:1998bm} and applied to quark-antiquark
operators \cite{Gockeler:1998ye,Gockeler:2010yr}.
Improved variants include the RI/SMOM and RI${}'$/SMOM schemes
\cite{Sturm:2009kb,Gorbahn:2010bf}, which differ in the way three-point
functions are treated.
In the RI/MOM and RI${}'$/MOM schemes, the subtraction is done at vanishing
operator momenta, which potentially generates additional sensitivity to
short-distance effects in the respective channel.
On the other hand, in the RI/SMOM and RI${}'$/SMOM schemes,
the subtraction of three-point functions is performed at the symmetric
Euclidean point, $-\mu^2$, by setting
\begin{align}
\label{eq:SMOM}
  p^2 = q^2 = (p+q)^2 = -\mu^2 \,, \qquad p\cdot q = \frac{\mu^2}{2} \,,
\end{align}
where the four-momenta $p$ and $q$ are as depicted in Fig.~1.
Thus, there is no channel with exceptional momenta in this scheme.
\begin{figure}[h]
\centerline{\includegraphics[width=0.3\textwidth]{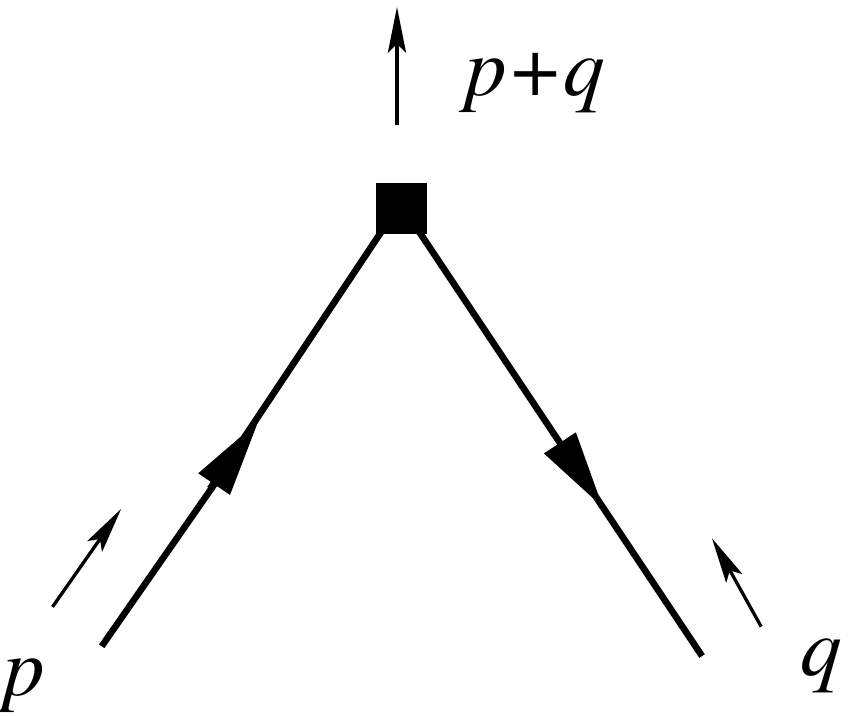}}
\caption{Matrix element $\langle \psi(q)\,O(-q-p)\,\bar{\psi}(p) \rangle$ of
bilinear quark operator in momentum space.
The black box denotes the operator and the solid lines the external quarks.}
\end{figure}

The next step after the nonpertubative renormalization is the perturbative
conversion of the results from one of the above schemes into the modified
minimal-subtraction ($\overline{\mathrm{MS}}$) scheme of dimensional
regularization, which serves as the worldwide standard in perturbative QCD
calculations.
Choosing the parameter $-\mu^2$ to be of the order of a few GeV${}^2$, such a
conversion can be done perturbatively order by order in the expansion in the
strong-coupling constant $\alpha_s(-\mu^2)$.
This matches the lattice simulations with the high-energy behavior determined
by conventional perturbation theory in the continuum using the 
$\overline{\mathrm{MS}}$ scheme.

The RI/SMOM to $\overline{\mathrm{MS}}$ conversion functions of non-singlet
bilinear quark operators without derivatives have been considered in
Refs.~\cite{Sturm:2009kb,Gracey:2010ci} at one loop and in
Refs.~\cite{Gorbahn:2010bf,Almeida:2010ns,Gracey:2011fb} at two loops.
In our previous paper \cite{Kniehl:2020sgo}, we extended this analysis to the
three-loop order numerically.
Our three-loop result for the (pseudo)scalar current has been confirmed by an
analytical calculation \cite{Bednyakov:2020ugu} in terms of constants
constructed earlier in Ref.~\cite{Kniehl:2017ikj}.

The corresponding conversions for the $n=2,3$ moments of the bilinear quark
operators of twist two with one or two covariant derivatives have been
considered in Refs.~\cite{Gracey:2010ci,Gracey:2011fb,Gracey:2011zn,Gracey:2011zg}
at the one- and two-loop orders.

In this paper, we extend this analysis to the three-loop order.
We concentrate on the cases of $n=2$ and $n=3$ and study the relevant operators
at the symmetric kinematical point up to three loops.
This paper is organized as follows.
In Section~2, we introduce our notations and definitions.
In Sections 3 and 4, we present our three-loop results for the $n=2$ and $n=3$
moments, respectively.
In Section~5, we conclude with a summary.

\section{Setup}

To fix the notation, we start from the following expression in Minkowski
coordinate space:
\begin{align}
  \int dx\,dy \,e^{-iq\cdot x-ip\cdot y} \langle \psi_{\xi,i}(x) O(0)
  \bar{\psi}_{\zeta,j}(y) \rangle =
  \delta_{ij} S_{\xi\xi'}(-q)
  \Lambda_{\xi'\zeta'}(p,q) S_{\zeta'\zeta}(p)\,,
\label{eq:matrix_element}
\end{align}
where $O$ stands for some bilinear quark operator, $\xi,\zeta$ are spinor
indices, $i,j$ are color indices in the fundamental representation of the
$SU(N)$ group, $S(q)$ is the quark propagator, and $\Lambda(p,q)$ is the
amputated Green's function, which is shown schematically in Fig.~1.

In the cases $n=2$ and $n=3$, we can write explicitly for any operators
$O_{\mu\nu}$ and $O_{\mu\nu\sigma}$:
\begin{align}
{\cal S} O_{\mu\nu} &= \frac{1}{2} \big(
   O_{\mu\nu} + O_{\nu\mu} \big) - \frac{1}{d}\, g_{\mu\nu} O_\alpha^\alpha \,, \\
{\cal S} O_{\mu\nu\sigma} &= \frac{1}{6} \big(
     O_{\mu\nu\sigma} + O_{\nu\mu\sigma} + O_{\nu\sigma\mu} + O_{\sigma\nu\mu}
   + O_{\sigma\mu\nu} + O_{\mu\sigma\nu} \big)  \nonumber \\
  &- \frac{1}{3(d+2)} \big( g_{\mu\nu} g_\sigma^\rho 
      + g_{\nu\sigma} g_\mu^\rho + g_{\sigma\mu} g_\nu^\rho \big)
      \big( O_{\rho\alpha\alpha'} + O_{\alpha\rho\alpha'} + O_{\alpha\alpha'\rho} \big)
      g^{\alpha\alpha'} \,,
\end{align}
where $g_{\mu\nu}$ is the metric tensor and $d=4-2\varepsilon$ is the
space-time dimension.

In the definition in Eq.~(\ref{eq:Odef}), we still have the freedom to define
in which directions the covariant derivatives act.
Thus, in the case with one derivative, we can define two operators,
\begin{align}
\label{eq:W2def}
{\cal S}O^{L}_{\mu\nu} = {\cal S}\bar{\psi} \gamma_\mu 
    \overset{\leftarrow}{D_\nu} \psi \,, \\
{\cal S}O^{R}_{\mu\nu} = {\cal S}\bar{\psi} \gamma_\mu 
     \overset{\rightarrow}{D_\nu} \psi \,, 
\end{align}
from which we can construct operators with either sign of charge
conjugation ($C$),
\begin{align}
  \label{eq:O2m1}
  O^{C=-1} = {\cal S}O^{L} + {\cal S}O^{R} \,,\\
  \label{eq:O2p1}
  O^{C=+1} = {\cal S}O^{L} - {\cal S}O^{R} \,,
\end{align}
where we have omitted the indices $\mu,\nu$ for the ease of notation.
Notice that the operators in Eqs.~(\ref{eq:O2m1}) and (\ref{eq:O2p1}) do not
mix under renormalization, so that the operator renomalization matrix is
diagonal in this basis.
In Refs.~\cite{Gracey:2010ci,Gracey:2011fb,Gracey:2011zn}, different operators,
called $W_2$ and $\partial W_2$, have been introduced.
These can be expressed in terms of the operators $O^L$ and $O^R$ with the help
of a suitable $2\times2$ transformation matrix, as
\begin{align}
\label{eq:W2transform}
 \frac{1}{2}
  \begin{pmatrix}
  W_2 \\
  \partial W_2
  \end{pmatrix} = 
  \begin{pmatrix}
  1\;\; & 0 \\
  1\;\; & 1
  \end{pmatrix}
  \begin{pmatrix}
  O^L \\
  O^R
  \end{pmatrix} \,.
\end{align}
The factor $1/2$ in Eq.~(\ref{eq:W2transform}) appears because it has
been omitted in the definitions of $W_2$ and $\partial W_2$ in
Refs.~\cite{Gracey:2010ci,Gracey:2011fb,Gracey:2011zn}.
We should also note that, in these papers, $W_2$ corresponds to the operator
where the covariant derivative acts to the right, while, according to our
definitions, the derivative in $W_2$ acts to the left.
Only with such conventions, we find agreement with
Refs.~\cite{Gracey:2010ci,Gracey:2011fb,Gracey:2011zn}.

For the operators with two derivatives, we introduce the following basis of
three operators:
\begin{align}
\label{eq:W3def}
{\cal S}O^{LL}_{\mu\nu\sigma} = {\cal S}\bar{\psi} \gamma_\mu 
    \overset{\leftarrow}{D_\nu} \overset{\leftarrow}{D_\sigma} \psi \,, \\
{\cal S}O^{LR}_{\mu\nu\sigma} = {\cal S}\bar{\psi} \gamma_\mu 
    \overset{\leftarrow}{D_\nu} \overset{\rightarrow}{D_\sigma} \psi \,, \\
\label{eq:W3def_end}
{\cal S}O^{RR}_{\mu\nu\sigma} = {\cal S}\bar{\psi} \gamma_\mu 
    \overset{\rightarrow}{D_\nu} \overset{\rightarrow}{D_\sigma} \psi \,.
\end{align}
From these operators, we can define the following combinations with definite
$C$ parities:
\begin{align}
  O_1^{C=-1} &= O^{LL} - 2O^{LR} + O^{RR} \,, \\
  O_2^{C=-1} &= O^{LL} + 2O^{LR} + O^{RR} \,, \\
  O_3^{C=+1} &= O^{LL} - O^{RR} \,, 
\end{align}
where we again omit the indices $\mu,\nu,\sigma$ for simplicity.
Operators $O_1$ and $O_2$ mix under renormalization, so that the $3\times3$
operator renormalization matrix takes a block diagonal form in this basis, with
one block of size $2\times2$ and one of size $1\times1$. 

In Refs.~\cite{Gracey:2010ci,Gracey:2011fb,Gracey:2011zg}, a different triplet
of operators, called $W_3$, $\partial W_3$, and $\partial\partial W_3$, has
been introduced.
We can express these in terms of the operators in
Eqs.~(\ref{eq:W3def})--(\ref{eq:W3def_end}) as
\begin{align}
  \begin{pmatrix}
  W_3 \\
  \partial W_3 \\
  \partial\partial W_3
  \end{pmatrix} = 
  \begin{pmatrix}
  1\;\; & 0\;\; & 0 \\
  1\;\; & 1\;\; & 0 \\
  1\;\; & 2\;\; & 1
  \end{pmatrix}
  \begin{pmatrix}
  O^{LL} \\
  O^{LR} \\
  O^{RR}
  \end{pmatrix} \,.
\end{align}
Similarly to the previous case, we find that the directions in which the
covariant derivatives act in the operator $W_3$ defined in
Refs.~\cite{Gracey:2010ci,Gracey:2011fb,Gracey:2011zg} should be flipped.
Upon this change, we find agreement with the previous one- and two-loop
calculations.

In order to renormalize the above operators, we use appropriate matrices $Z$ of
renormalization constants, a $2\times2$ matrix for $n=2$ and a $3\times3$ matrix
for $n=3$.
In the $\overline{\mathrm{MS}}$ scheme, we can write
\begin{align}
  Z &= 1 + \frac{Z_1}{\varepsilon} + \frac{Z_2}{\varepsilon^2} + \frac{Z_3}{\varepsilon^3} + \cdots\,,
\end{align}
where $Z_i$ are constant matrices depending on the QCD coupling constant,
\begin{align}
   a &= \frac{\alpha_s}{4\pi} \,.
\end{align}
These matrices can be related to the matrix of anomalous dimensions $\gamma$ by
the following matrix equations:
\begin{align}
a \,\partial_a Z_1 &= - \gamma \,, \\
a \,\partial_a Z_2 &= a \,\partial_a \left( \frac{1}{2} Z_1^2 \right)
    + \beta \,\partial_a Z_1 - \xi \gamma_3 \,\partial_\xi Z_1 \,,  \\
a \,\partial_a Z_3 &= a \,\partial_a \left( \frac{Z_1 Z_2 + Z_2 Z_1}{2} - \frac{1}{3} Z_1^3 \right)
  + \beta \,\partial_a \left( Z_2 - \frac{1}{2} Z_1^2 \right)
  \nonumber\\
   &- \xi \gamma_3 \,\partial_\xi \left( Z_2 - \frac{1}{2} Z_1^2 \right)  \,,
\end{align}
where $\beta$ is QCD $\beta$ function, $\xi$ is the gauge parameter, and
$\gamma_3$ is the anomalous dimension associated with the latter
\cite{Larin:1993tp}. 

The matrix $\gamma$ for $n=2$ has been evaluated analytically through $O(a^3)$
in Ref.~\cite{Gracey:2011zn}.
The corresponding matrix for $n=3$ can be found in
Ref.~\cite{Gracey:2011zg}.\footnote{%
Notice that the definition of $\gamma$ given by Eq.~(2.8) in
Ref.~\cite{Gracey:2011zg} differs by a factor of 2 from the definition used for
the results in Eq.~(2.10) therein.}
Moreover, in Ref.~\cite{Gracey:2011zg}, the nondiagonal matrix elements are
only given through order $O(a^2)$.
We evaluate the missing $O(a^3)$ contributions numerically for the color group
SU(3).
In the basis $(W_3,\partial W_3,\partial\partial W_3)$, Eq.~(2.10) in
Ref.~\cite{Gracey:2011zg} should be extended by the following
three-loop contributions
\begin{align}
\gamma_{12}^{W_3,O(a^3)}(a) &= a^3 \big(-385.466 + 66.199 n_f + 0.5329 n_f^2 \big) \,,\\
\gamma_{13}^{W_3,O(a^3)}(a) &= a^3 \big(-170.641 + 24.822 n_f + 0.3107 n_f^2 \big)\,,
\end{align}
where $n_f$ is the number of light-quark flavors.

To represent our results, we adopt the tensor decompositions from
Refs.~\cite{Gracey:2011zn,Gracey:2011zg}.
It is convenient to contract the open indices of the operators $O_{\mu\nu}$ and
$O_{\mu\nu\sigma}$ with the light-cone vector $\Delta$, with  $\Delta^2=0$.
This automatically takes into account the symmetry and the tracelessness of the
operators.
Specifically, we write
\begin{align}
\label{eq:lambda2}
\frac{1}{-i}\Lambda_2(p,q) &=  \slaDelta \Bigg[
      2(p\cdot\Delta) \,F_1
    + 2(q\cdot\Delta) \,F_2 \Bigg]
\nonumber\\
 &+ \frac{1}{\mu^2} \slap \Bigg[
      (p\cdot\Delta)^2 \,F_3
    + 2(p\cdot\Delta)(q\cdot\Delta) \,F_4
    +  (q\cdot\Delta)^2 \,F_5 \Bigg]
\nonumber\\
 &+ \frac{1}{\mu^2} \slaq \Bigg[
      (p\cdot\Delta)^2 \,F_6 
    + 2(p\cdot\Delta)(q\cdot\Delta) \,F_7
    +  (q\cdot\Delta)^2 \,F_8 \Bigg]
\nonumber\\
 &+ \frac{1}{\mu^2} \Gamma_{3,\Delta pq} \Bigg[
      2(p\cdot\Delta) \,F_{9}
    + 2(q\cdot\Delta) \,F_{10} \Bigg]\,,
\\
\label{eq:lambda3}
\frac{1}{(-i)^2\mu^2}\Lambda_3(p,q) &=  \frac{1}{\mu^2} \slaDelta \Bigg[
      3 (p\cdot\Delta)^2 \,F_1
    + 6 (p\cdot\Delta)(q\cdot\Delta) \,F_2
    + 3 (q\cdot\Delta)^2 \,F_3 \Bigg]
\nonumber\\
 &+ \frac{1}{\mu^4} \slap \Bigg[
      (p\cdot\Delta)^3 \,F_4
    + 3(p\cdot\Delta)^2(q\cdot\Delta) \,F_5 
    + 3(p\cdot\Delta)(q\cdot\Delta)^2 \,F_6 
    + (q\cdot\Delta)^3 \,F_7 \Bigg]
\nonumber\\
 &+ \frac{1}{\mu^4} \slaq \Bigg[
      (p\cdot\Delta)^3 \,F_8
    + 3(p\cdot\Delta)^2(q\cdot\Delta) \,F_9 
    + 3(p\cdot\Delta)(q\cdot\Delta)^2 \,F_{10} 
    + (q\cdot\Delta)^3 \,F_{11} \Bigg]
\nonumber\\
 &+ \frac{1}{\mu^4} \Gamma_{3,\Delta pq} \Bigg[
      3(p\cdot\Delta)^2 \,F_{12}
    + 6(p\cdot\Delta)(q\cdot\Delta) \,F_{13} 
    + 3(q\cdot\Delta)^2 \,F_{14} \Bigg] \,.
\end{align}
Here 
\begin{align}
\Gamma_{3,\mu\nu\sigma} = \frac{1}{3!}  \gamma_{[\mu}\gamma_\nu\gamma_{\sigma]} 
\end{align}
is the fully antisymmetric combination of the Dirac $\gamma$ matrices, and we
use the short-hand notation
$\Gamma_{3,\Delta pq}=\Gamma_{3,\mu\nu\sigma}\Delta^\mu p^\nu q^\sigma$.
With the definitions in Eqs.~(\ref{eq:lambda2}) and (\ref{eq:lambda3}), the
definitions of the formfactors $F_1,\dots,F_{10}$ and $F_1,\dots,F_{14}$
coincide with those in Ref.~\cite{Gracey:2011zn} and \cite{Gracey:2011zg},
respectively.

We refrain from describing our calculation because it is similar to the one in
Ref.~\cite{Kniehl:2020sgo}, where details may be found, and merely list our
results, which we do for the $n=2$ case in Section~3 and for the $n=3$ case
in Section~4.

\section{Numerical results for $n=2$ moment}

Here, we present the numerical results for the formfactors $F^{L}_j$ and
$F^{R}_j$ of the $n=2$ moment at three loops in the $\overline{\mathrm{MS}}$
scheme.
For $F^{L}_j$, we have
\begin{align}
F^L_1 &=  a \big( 0.87497670537933942370 \big) \nonumber\\
 &+ a^2 \big( 18.69246420435249196 - 2.5121840774766979282 n_f \big) \nonumber\\
 &+ a^3 \big( 767.149(2) - 147.4921(2) n_f + 3.99834(1) n_f^2 \big) \,, \\
F^L_2 &=  0.5 + a \big( -1.6874417634483485593 \big) \nonumber\\
 &+ a^2 \big( -21.75488024301858658 + 2.394054755622383881 n_f \big) \nonumber\\
 &+ a^3 \big( -624.064(4) + 123.5347(3) n_f - 2.81311(1) n_f^2 \big) \,, \\ 
F^L_3 &= a \big( -0.62655873962365026074 \big) \nonumber\\
 &+ a^2 \big( -7.8618849118581104 + 0.4799770959727058987 n_f \big) \nonumber\\
 &+ a^3 \big( -254.42(1) + 48.002(1) n_f - 0.71738(1) n_f^2 \big) \,, \\
F^L_4 &= a \big( -0.82709837483873229037 \big) \nonumber\\
 &+ a^2 \big( -8.42794691505549605 + 0.8725648296451055402 n_f \big) \nonumber\\
 &+ a^3 \big( -313.419(4) + 64.3005(4) n_f - 1.38245(3) n_f^2 \big) \,, \\
F^L_5 &= a \big( -1.3116507463261931407 \big) \nonumber\\
 &+ a^2 \big( -30.32774293571586111 + 2.4134870684309379409 n_f \big) \nonumber\\
 &+ a^3 \big( -1281.20(1) + 213.0615(9) n_f - 5.33964(3) n_f^2 \big) \,, \\
F^L_6 &=  a \big( -0.4661270314515846370 \big) \nonumber\\
 &+ a^2 \big( -14.05284249471245715 + 0.4507104624332595895 n_f \big) \nonumber\\
 &+ a^3 \big(  -467.649(5) + 63.8280(5) n_f - 0.71048(3) n_f^2 \big) \,, \\
F^L_7 &= a \big( -0.79783174129928598074 \big) \nonumber\\
 &+ a^2 \big( -23.16109121325045065 + 1.2245098599795996977 n_f \big) \nonumber\\
 &+ a^3 \big( -890.390(6) + 134.2270(5) n_f - 2.708 n_f^2 \big) \,, \\
F^L_8 &= a \big( -0.8455237148746085037 \big) \nonumber\\
 &+ a^2 \big( -10.93560591432546472 + 0.849974752412507047 n_f \big) \nonumber\\
 &+ a^3 \big( -404.35(2) + 72.164(2) n_f - 1.41340(7) n_f^2 \big) \,, \\
F^L_9 &= a \big( 0.22222222222222222222 \big) \nonumber\\
 &+ a^2 \big( 4.41926247296556700 - 0.22143896997362032903 n_f \big) \nonumber\\
 &+ a^3 \big( 170.416(2) - 25.7049(1) n_f + 0.355738(1) n_f^2 \big) \,, \\
F^L_{10} & = a \big( 0.8195143283064261733 \big) \nonumber\\
 &+ a^2 \big( 15.47442232012187938 - 1.2832916030122051316 n_f \big) \nonumber\\
 &+ a^3 \big( 580.239(3) - 108.8446(4) n_f + 2.83040(3) n_f^2 \big) \,.
\end{align}

Via crossing symmetry in the decomposition in Eq.~(\ref{eq:lambda2}), we obtain
for $F^R_j$
\begin{align}
F^R_{1,2} = F^L_{2,1}\,, \qquad
F^R_{3,4,5} = F^L_{8,7,6}\,, \qquad
F^R_{6,7,8} = F^L_{5,4,3}\,, \qquad
F^R_{9,10} = F^L_{10,9}\,.
\end{align}

Comparing with the previous calculations by Gracey
\cite{Gracey:2010ci,Gracey:2011fb,Gracey:2011zn}, we find agreement by
verifying the relations $F^L_j=-\frac{1}{2} \Sigma^{W_2}_{(j)}$ and
$F^L_j+F^R_j=-\frac{1}{2} \Sigma^{\partial W_2}_{(j)}$ for $j=1,\ldots,10$
through the two-loop order.

\section{Numerical results for $n=3$ moment}

Here, we present the numerical results for the formfactors $F^{LL}_j$,
$F^{RR}_j$, and $F^{LR}_j$ of the $n=3$ moment at three loops in the
$\overline{\mathrm{MS}}$ scheme.
For $F^{LL}_j$, we have
\begin{align}
F^{LL}_1 &= 
 a \big(     0.12809418462663994519 \big) \nonumber\\
 &+a^2 \big(     3.57396324725023741 - 0.3927663257641307273 n_f \big) \nonumber\\
 &+a^3 \big(     142.934(4) - 23.3744(5) n_f + 0.38322(1) n_f^2 \big) \,, \\
F^{LL}_2 &= 
 a \big(     0.64814814814814814815 \big) \nonumber\\
 &+a^2 \big(     11.92146760129898963 - 1.6685976202307340604 n_f \big) \nonumber\\
 &+a^3 \big(     470.434(6) - 94.2894(8) n_f + 2.69132(1) n_f^2 \big) \,, \\
F^{LL}_3 &=  0.33333333333333333333 +
 a \big(     - 1.5801847945918187101 \big) \nonumber\\
 &+a^2 \big(     - 23.39093305099714828 + 2.878059562968590479 n_f \big) \nonumber\\
 &+a^3 \big(     - 784.543(3) + 157.3102(3) n_f - 4.15775(1) n_f^2 \big) \,, \\
F^{LL}_4 &= 
 a \big(     - 0.34254600335127144000 \big) \nonumber\\
 &+a^2 \big(     - 4.53423910035660995 + 0.2361911259199855009 n_f \big) \nonumber\\
 &+a^3 \big(     - 138.88(5) + 25.985(4) n_f - 0.34336(2) n_f^2 \big) \,, \\
F^{LL}_5 &= 
 a \big(     - 0.42384760642772557827 \big) \nonumber\\
 &+a^2 \big(     - 4.59354266387785339 + 0.3660534795383755215 n_f \big) \nonumber\\
 &+a^3 \big(     - 152.53(5) + 31.366(5) n_f - 0.528634(1) n_f^2 \big) \,, \\
F^{LL}_6 &= 
 a \big(     - 0.5842793145997912020 \big) \nonumber\\
 &+a^2 \big(     - 4.85747463745270533 + 0.6841872158771917737 n_f \big) \nonumber\\
 &+a^3 \big(     - 193.46(4) + 43.965(3) n_f - 1.08542(4) n_f^2 \big) \,, \\
F^{LL}_7 &= 
 a \big(     - 1.0677459370968307259 \big) \nonumber\\
 &+a^2 \big(     - 23.37760835366161187 + 2.126160364999081326 n_f \big) \nonumber\\
 &+a^3 \big(     - 1037.27(2) + 177.931(2) n_f - 4.87027(7) n_f^2 \big) \,, \\
F^{LL}_8 &= 
 a \big(     - 0.22222222222222222222 \big) \nonumber\\
 &+a^2 \big(     - 7.10270791265820791 + 0.1633837590014029747 n_f \big) \nonumber\\
 &+a^3 \big(     - 223.72(4) + 28.698(4) n_f - 0.241101(1) n_f^2 \big) \,, \\
F^{LL}_9 &= 
 a \big(     - 0.28292698728195749432 \big) \nonumber\\
 &+a^2 \big(     - 9.25460402258154041 + 0.2645583674342950950 n_f \big) \nonumber\\
 &+a^3 \big(     - 306.93(4) + 40.838(5) n_f - 0.426065 n_f^2 \big) \,, \\
F^{LL}_{10} &= 
 a \big(     - 0.47732484248508078617 \big) \nonumber\\
 &+a^2 \big(     - 15.44021253016360790 + 0.7240160519079720091 n_f \big) \nonumber\\
 &+a^3 \big(     - 587.16(3) + 86.038(3) n_f - 1.64434(2) n_f^2 \big) \,, \\
F^{LL}_{11} &= 
 a \big(     - 0.5615109786022296830 \big) \nonumber\\
 &+a^2 \big(     - 7.60796010282396424 + 0.606188782359786649 n_f \big) \nonumber\\
 &+a^3 \big(     - 288.83(1) + 50.148(2) n_f - 1.03938(10) n_f^2 \big) \,, \\
F^{LL}_{12} &= 
 a \big(     0.060704765059735272099 \big) \nonumber\\
 &+a^2 \big(     1.27591955027728723 - 0.04990488421511731147 n_f \big) \nonumber\\
 &+a^3 \big(     49.32(1) - 6.8348(9) n_f + 0.07301(1) n_f^2 \big) \,, \\
F^{LL}_{13} &= 
 a \big(     0.12140953011947054420 \big) \nonumber\\
 &+a^2 \big(     2.55370383484854647 - 0.12707055462464682228 n_f \big) \nonumber\\
 &+a^3 \big(     98.40(1) - 14.3840(9) n_f + 0.206262(1) n_f^2 \big) \,, \\
F^{LL}_{14} &= 
 a \big(     0.45889950244920457284 \big) \nonumber\\
 &+a^2 \big(     8.64602611504816215 - 0.7578066395741738465 n_f \big) \nonumber\\
 &+a^3 \big(     322.542(7) - 62.2614(7) n_f + 1.72278(2) n_f^2 \big) \,.
\end{align}

Via crossing symmetry in the decomposition in Eq.~(\ref{eq:lambda3}), we obtain
for $F^{RR}_j$
\begin{align}
F^{RR}_{1,2,3} = F^{LL}_{3,2,1}\,, \quad
F^{RR}_{4,5,6,7} = F^{LL}_{11,10,9,8}\,, \quad
F^{RR}_{8,9,10,11} = F^{LL}_{7,6,5,4}\,, \quad
F^{RR}_{12,13,14} = F^{LL}_{14,13,12}\,.
\end{align}

Finally, for $F^{LR}_j$ we have
\begin{align}
F^{LR}_1 &= F^{LR}_3 =
 a \big(     0.45522361895958633728 \big) \nonumber\\
 &+a^2 \big(     8.88767955565142389 - 1.2820230592203345582 n_f \big) \nonumber\\
 &+a^3 \big(     368.500(5) - 74.9538(4) n_f + 2.28234(1) n_f^2 \big) \,, \\
F^{LR}_2 &=  0.16666666666666666667
  + a \big(     - 0.91896983417115119333 \big) \nonumber\\
 &+a^2 \big(     - 12.94227294752102117 + 1.6292211796126293778 n_f \big) \nonumber\\
 &+a^3 \big(     - 422.739(7) + 86.3035(6) n_f - 2.29625(1) n_f^2 \big) \,, \\
F^{LR}_4 &= F^{LR}_{11} = 
 a \big(     - 0.28401273627237882074 \big) \nonumber\\
 &+a^2 \big(     - 3.32764581150150047 + 0.2437859700527203978 n_f \big) \nonumber\\
 &+a^3 \big(     - 115.52(3) + 22.016(3) n_f - 0.37402(3) n_f^2 \big) \,, \\
F^{LR}_5 &= F^{LR}_{10} =
 a \big(     - 0.33640422333931270222 \big) \nonumber\\
 &+a^2 \big(     - 3.64571691677851411 + 0.3756487722159301382 n_f \big) \nonumber\\
 &+a^3 \big(     - 141.21(5) + 27.501(4) n_f - 0.632127(1) n_f^2 \big) \,, \\
F^{LR}_6 &= F^{LR}_9 =
 a \big(     - 0.40433651740142803852 \big) \nonumber\\
 &+a^2 \big(     - 10.87040428448957907 + 0.70201836002985790003 n_f \big) \nonumber\\
 &+a^3 \big(     - 442.55(5) + 69.922(4) n_f - 1.61609(1) n_f^2 \big) \,, \\
F^{LR}_7 &= F^{LR}_8 = 
 a \big(     - 0.2439048092293624148 \big) \nonumber\\
 &+a^2 \big(     - 6.95013458205424924 + 0.2873267034318566148 n_f \big) \nonumber\\
 &+a^3 \big(     - 243.93(3) + 35.130(3) n_f - 0.46937(3) n_f^2 \big) \,, \\
F^{LR}_{12} &= F^{LR}_{14} = 
 a \big(     0.087443383088412876049 \big) \nonumber\\
 &+a^2 \big(     1.67025543169975743 - 0.09772109576729624122 n_f \big) \nonumber\\
 &+a^3 \big(     64.285(8) - 10.3017(10) n_f + 0.164148(1) n_f^2 \big) \,, \\
F^{LR}_{13} &= 
 a \big(     0.22583598672341225432 \big) \nonumber\\
 &+a^2 \big(     4.07752442951393566 - 0.37450630303729499795 n_f \big) \nonumber\\
 &+a^3 \big(     151.82(1) - 30.4658(9) n_f + 0.855783(1) n_f^2 \big) \,.
\end{align}

Comparing with the previous calculations by Gracey
\cite{Gracey:2010ci,Gracey:2011fb,Gracey:2011zg}, we find agreement by
verifying the relations $F^{LL}_j=-\Sigma^{W_3}_{(j)}$,
$F^{LL}_j+F^{LR}_j=-\Sigma^{\partial W_3}_{(j)}$, and
$F^{LL}_j+2F^{LR}_j+F^{RR}_j=-\Sigma^{\partial\partial W_3}_{(j)}$
for $j=1,\ldots,14$ through the two-loop order.

\section{Conclusion}

In this paper, we have calculated the $n=2$ and $n=3$ moments of the
twist-two non-singlet bilinear quark operators in SMOM kinematics
at three loops in QCD.
This allows us to match, with unprecedented precision, lattice QCD simulations
of these quantities to their high-energy behaviors in the continuum limit as
determined from perturbative QCD calculations in the $\overline{\mathrm{MS}}$
scheme.
We have presented the relevant conversion factors between the RI/SMOM and
$\overline{\rm MS}$ schemes in numerical form, ready to be used by the lattice
community.
The three-loop corrections are comparable in size to the two-loop
contributions available from
Refs.~\cite{Gracey:2010ci,Gracey:2011fb,Gracey:2011zn,Gracey:2011zg}, which we
were able to reproduce after clarifying some issues with the definitions.

\section*{Acknowledgments}

This work was supported in part by DFG Research Unit FOR~2926 through Grant
No.\ 409651613.
O.L.V. is grateful to V.~Braun and M.~G\"ockeler for fruitful discussions
and to the University of Hamburg for the warm hospitality.



\end{document}